\documentclass[10pt,preprint2]{aastex}
\usepackage{color}
\usepackage{ulem}

\newcommand{\fig}[1]{Fig.~\ref{#1}}
\newcommand{\eq}[1]{eq.~(\ref{#1})}
\newcommand{\etal}{\textit{et al.}}
\newcommand{\indwater}{_{\scriptsize\textrm{H}_2\textrm{O}}}
\newcommand{\indih}{_{\scriptsize\textrm{H}_2}}
\newcommand{\indice}[1]{_{\scriptsize\textrm{#1}}}

\shorttitle{H2-H2O mixtures}
\shortauthors{Soubiran et al.}

\begin{document}

\title{Miscibility calculations for water and hydrogen in giant planets}

\author{Fran\c{c}ois Soubiran\altaffilmark{1} and Burkhard
Militzer\altaffilmark{1,2}}
\affil{$^1$Department of Earth and Planetary Science, University of California,
    Berkeley, CA 94720, United States}
\affil{$^2$Department of Astronomy, University of California,
    Berkeley, CA 94720, United States}

\author{Apr. 22, 2015\\ \textit{Manuscript accepted for publication in} The Astrophysical Journal.}

\begin{abstract}

  We present results from \textit{ab initio} simulations of liquid
  water-hydrogen mixtures in the range from 2 to 70~GPa and from 1000
  to 6000~K, covering conditions in the interiors of ice giant planets
  and parts of the outer envelope of gas giant planets. In addition
  to computing the pressure and the internal energy, we derive the
  Gibbs free energy by performing a thermodynamic integration. For
  all conditions under consideration, our simulations predict hydrogen
  and water to mix in all proportions. The thermodynamic behavior of
  the mixture can be well described with an ideal mixing
  approximation. We suggest a substantial fraction of water and
  hydrogen in giant planets may occur in homogeneously mixed form
  rather than in separate layers. The extend of mixing depends on the
  planet's interior dynamics and its conditions of formation, in
  particular on how much hydrogen was present when icy planetesimals
  were delivered.  Based on our results, we do not predict water-hydrogen 
 mixtures to phase separate during any stage of the evolution of giant planets. We also 
 show that the hydrogen content of an exoplanet is
  much higher if the mixed interior is assumed.

\end{abstract}


\section{Introduction}

The past decade has been a period of extraordinary discoveries in the
field of exoplanets~\citep{exoarch}. For the first time, we now have
quantitative estimates for occurrence rates of different types of
planets in our galaxy. A recent study by \citet{petigura_2013} that focused on planets 
with periods up to 100
days in the Kepler sample demonstrates the prevalence of Neptune and
sub-Neptune exoplanets. In this sample, the planets with a radius
larger than 1.5 Earth radii have a mean density~\citep{weiss_2014}
that implies that they are composed of both heavy elements (rocks and ices) as
well as gas (hydrogen and helium) (for mass-radius relation, see
\citet{seager_2007}). Despite the absence of sub-Neptune planets in
our solar system, we may be able to place constraints on the formation
process of ice giants by studying the interior and the evolution of Uranus
and Neptune. Both planets have sizable gaseous envelopes surrounding cores
composed of heavier elements.

The core accretion model~\citep{pollack_1996, helled_2014, bodenheimer_2014} suggests that 
the giant planets form in two distinct phases. First rocky and icy planetesimals 
accumulate to form a dense core. Once a critical core mass has been reached, a runaway 
accretion of hydrogen-helium gas sets in and lasts until all gas from the planet's 
neighborhood has been depleted. Most often it has been assumed that very little gas is 
present when the initial core forms~\citep{pollack_1996}. Because of this assumption, 
planetary interior models typically assume a dense core of rock and ice with a sharp 
transition to a gaseous outer envelope that is composed of hydrogen, helium, and a small 
fraction of heavier elements~ \citep{stevenson_1985, 
hubbard_1995,hubbard_1999,guillot_1999,guillot_2005,militzer_2008}. While the 
heavier element fraction in Jupiter's envelope has been measured {\it in situ} by the 
Galileo entry probe \citep{wong_2004}, there is no direct measurement that characterizes 
the state of a giant planet core. It is thus plausible that some mixing of gas and icy 
planetesimals has occurred when a giant planet formed and that it may have persisted until 
today.

The mean density of Uranus and Neptune, on the other hand,
  suggests that they have a much higher fraction of heavy elements
   than Jupiter and Saturn \citep[pp.
  282-295]{hubbard_1984}. Based on cosmological abundances, it is
  assumed that water is among the dominant species, followed by
  methane and amonia, even though only trace amounts of it have been
detected spectroscopically in the atmospheres of Uranus and Neptune
\citep{encrenaz_2003}. This paper focuses on water-hydrogen mixtures
because of the prevalence of water in our solar system and the
possibility that gases and ices mix when giant planets form.

A recent study \citep{fortney_2013} focused on the late accretion of
planetesimals when a large, gaseous envelope is present. It was
predicted that even large planetesimals of 100 km in diameter may
disintegrate upon entry and the material would be distributed
throughout the gas envelope. Depending on the mixing properties of
hydrogen and heavier elements, this material may either remain in
the envelope or gradually settle onto the existing core. We will
study hydrogen-water mixture in this article because little
information is available about how well hydrogen mixes with other elements
at high pressure. The phase separation of hydrogen and helium has been
predicted to occur in the interiors of
Saturn~\citep{stevenson_1977,fortney_2004,morales_2009,soubiran_2013} and also of
Jupiter~\citep{WilsonMilitzer2010}. The resulting release of
gravitational energy has been named as a reason for the observed excess in
Saturn's luminosity. It would be of interest to know whether a similar
process involving the separation of water-hydrogen mixtures could
operate in the interiors of ice giant planets. If a phase separation
occurs, the envelope would be progressively depleted of heavy
elements, which affects the interior structure and the luminosity of a
giant planet.

On the contrary, it is possible that the core of an initially
differentiated planet would dissolve into the surrounding layer of
hot, dense hydrogen. Results from recent {\it ab initio} simulations
predict that the rocky and icy components of the cores in Jupiter and
Saturn are miscible in metallic
hydrogen~\citep{WilsonMilitzer2012,wahl_2013, gonzalez_2014}. While a
mixture state is thermodynamically preferred, it is not known how
fast the cores of giant planets erode because gravitational forces
counteract the advection of heavy elements. This may give rise to a
semi-convective regime that has already been suggested to occur in the
gas giant interiors \citep{leconte_2012,leconte_2013}. 

Using recent updates of the gravitational moments of Uranus and
Neptune \citep{jacobson_2007, jacobson_2009}, revised interior models
have been constructed. \citet{helled_2011} fitted a density profile to
these data and showed that they could be satisfied by a single layer
of hydrogen and helium with a compositional gradient of heavier
elements. On the other hand, \citet{nettelmann_2013} matched the
available constraints by constructing a typical three layers model
assuming a rocky core, surrounded by an intermediate layer of water in a
liquid or superionic state \citep{cavazzoni_1999, french_2009a,
  wilson_2013}, and a gaseous outer envelope. While Helled's model
assumes water and hydrogen are completely miscible throughout the
planet interior, Nettelmann \etal~conversely assume both fluids
would not mix at the interface of the two layers at approximately
10~GPa and 2000~K. This underlines why the mixing properties of water
and hydrogen are important for giant planet interiors.

Both aforementioned models assume an adiabatic, fully convective
behavior for each layer. This assumption is not consistent with the
observed luminosity of Uranus~\citep{pearl_1990, pearl_1991}.
\citet{podolak_1990} suggested a stably stratified interior model
instead.  Nevertheless, a convective layer of a conducting material is
needed to sustain a magnetic dynamo and to produce the quadrupolar
fields observed for Uranus and Neptune
\citep{ness_1986,ness_1989,connerney_1987,connerney_1991}. Water is
assumed to be the dominant species in the layer where magnetic fields
of ice giant planets are generated. If water and hydrogen mix, a
gradient of composition may be introduced into the planet's interior
during formation. It would thus be possible for semi-convection to
operate in ice giant planets which would reduce the overall heat flux
and allow for a vigorous convection in some layers.

Some experimental data for water-hydrogen mixtures are available. 
\citet{seward_1981} studied mixtures up to 0.25~GPa and observed a phase 
separation below 650~K. Based on this result, the presence of water clouds in 
the deep atmosphere of the ice giants has been inferred 
\citep{fegley_1986}. More recently, \citet{bali_2013} investigated the 
properties of water-hydrogen in the presence of minerals under deep-Earth 
conditions up to 2.5~GPa. Hydrogen was released due to chemical reactions 
between water and the minerals. The analysis of micro-inclusions of fluid 
within the minerals showed that phase separation of hydrogen and water was 
possible at temperatures below 1200~K. These findings favor differentiated 
interior models for Uranus and Neptune. 

At the present time, no experimental data are available for higher pressure 
and temperature but materials under such conditions can be studied 
efficiently with the \textit{ab initio} simulations that we used throughout 
this paper. Results from such simulations have been shown to agree very 
well with shock wave measurements for 
hydrogen~\citep{lenosky_1997,knudson_2004,loubeyre_2012}. Similarly results 
of shock wave experiments of water have been matched with \textit{ab 
initio} simulations~\citep{knudson_2012}. Recently improvements have been 
made to compute the dielectric constant of water with such 
methods~\citep{pan_2013}. Here we use the same technique to study 
water-hydrogen mixtures from 2~GPa to 70~GPa and from 1000~K to 6000~K. For 
various mixing ratios, we compute the equation of state, in particular the 
pressure and internal energy as function of density and temperature. In 
addition we derive the Gibbs free energy of mixing by performing a 
thermodynamic integration. We show that the mixtures behave close to an 
ideal mixture. The entropy of mixing is the dominant term in the Gibbs free 
energy of mixing indicating that no phase separation occurs at all 
conditions under consideration.

\begin{figure}[!ht]
\centering
\includegraphics[width=\columnwidth]{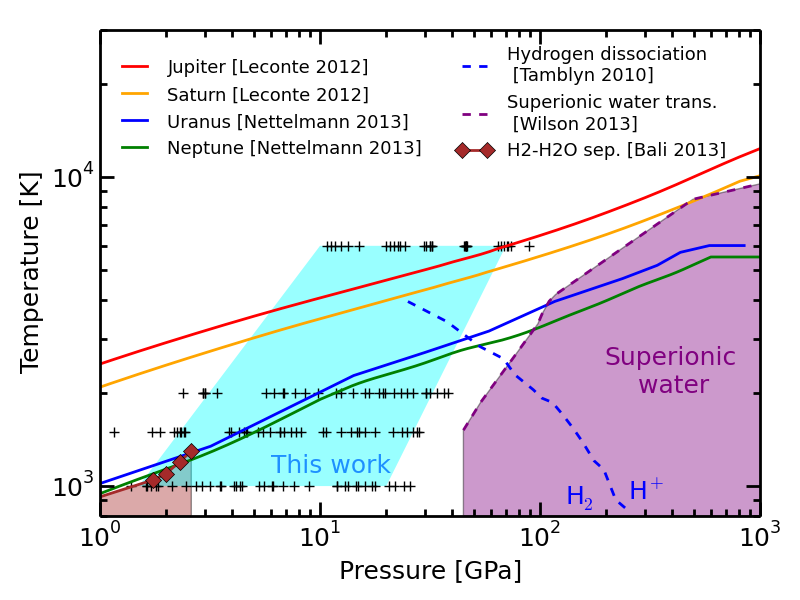}
\caption{\label{fig:phase_diag} Pressure-temperature diagram with the predicted interior 
profiles for the solar giant planets. 
The different expected phase transitions are also plotted. The light brown region is where 
Bali \etal~found a phase separated system. 
The cyan region shows the parameter range studied in this work. There, no phase 
separation was found. The plus symbols mark specific simulation conditions.}
\end{figure}

We also analyze changes in ionic species that are present in the mixture
as a function of the pressure and the temperature. Finally we comment on some
implications of computed miscibility properties. We show for instance
that a sharp transition from a hydrogen-rich envelope to a water-rich
phase is thermodynamically unstable and that a three layer picture for
the icy giants is a simplification. Furthermore we compare the
mass-radius relationship for Neptune-like and sub-Neptune planets
depending on their differentiation. We found that for given radius and
mass, a fully differentiated planet has a much lower hydrogen content
than a homogeneously mixed planet.

\section{Simulation methods}
Our investigations of the hydrogen-water mixture rely on \textit{ab
  initio} molecular dynamics (MD) simulations in which the nuclei are
treated as classical particles while the electrons are considered
quantum mechanically using density functional theory (DFT)
\citep{hohenberg_1964}. We used the Vienna Ab initio Simulation
Package (VASP) \citep{kresse_1996}. The time step of the MD
simulations was set to 0.2~fs, which is short enough to describe the
molecular vibrations accurately. Each simulation was performed for a
minimum of 0.5~ps and up to 7~ps for the lowest temperatures. To keep
the temperature constant, we employed a Nos\'e thermostat
\citep{nose_1984, nose_1991}. For the electrons, we used a Fermi-Dirac
distribution within a finite temperature scheme \citep{mermin_1965}.
We used projector augmented wave (PAW) pseudopotentials
\citep{blochl_1994} with a cut-off radius of $r_\textrm{cut}=0.8~a_0$
for hydrogen and $r_\textrm{cut}=1.1~a_0$ for oxygen. We used a
plane-wave basis energy cut-off at $1100$~eV. To sample the Brillouin
zone, we used the Baldereschi point \citep{baldereschi_1973}, except
for pure water, for which a $2\times2\times2$ Monkhorst-Pack K-point
grid \citep{monkhorst_1976} was used. We performed a few simulations
with finer K-point grids and found consistent results within the
statistical errorbars.  For the exchange-correlation functional, we
chose the generalized gradient approximation (GGA) of the Perdew,
Burke \& Ernzerhof (PBE) type \citep{perdew_1996} because it gave
reasonable results for pure hydrogen
\citep{caillabet_2011,loubeyre_2012} and pure water
\citep{french_2009b,knudson_2012}.  We also verified our predictions by performing
  additional simulations with the van der Waals density functional
  (vdw-DF) by \citet{dion_2004,klimes_2011}. The resulting mixing
  properties were in agreement with our PBE predictions at 1000~K. A
  more detailed comparison of the PBE and van der Waals functionals is
  given in the recent work by \citet{santra_2013}.

To explore the phase separation at a given pressure and temperature, it is
necessary to determine the Gibbs free energy. Standard molecular
dynamics only provide the internal energy and the pressure but not the
entropy of the system. Therefore we performed a thermodynamic integration using
an auxiliary classical pair potential
\citep{wijs_1998, morales_2009,WilsonMilitzer2010,WilsonMilitzer2012,WilsonMilitzer2012b,
McMahon2012,wahl_2013} because it provides the Helmholtz free energy difference
$F_1-F_0$ between two systems characterized by two different potentials
$U_0(\{\bold{r}\})$ and $U_1(\{\bold{r}\})$:

\begin{equation}\label{eq:tdi}
F_1-F_0=\int_0^1 \langle U_1-U_0 \rangle_\lambda \textrm{d}\lambda,
\end{equation}

where the parameter $\lambda$ defines a hybrid potential
$U_\lambda=U_0+\lambda (U_1 + U_0)$. The $\langle.\rangle_\lambda$
means that we take the average over a trajectory computed with the
potential $U_\lambda$. We performed the integration in two distinct
steps. First we integrated between the potential as given by the DFT,
$U\indice{DFT}$, and a set of classical pair potentials,
$U\indice{cl}$, that we constructed by matching the forces on
 configurations taken from a DFT trajectory that was computed beforehand for a
particular temperature, density, and composition. 

\begin{figure}[!ht]
\centering
\includegraphics[width=\columnwidth]{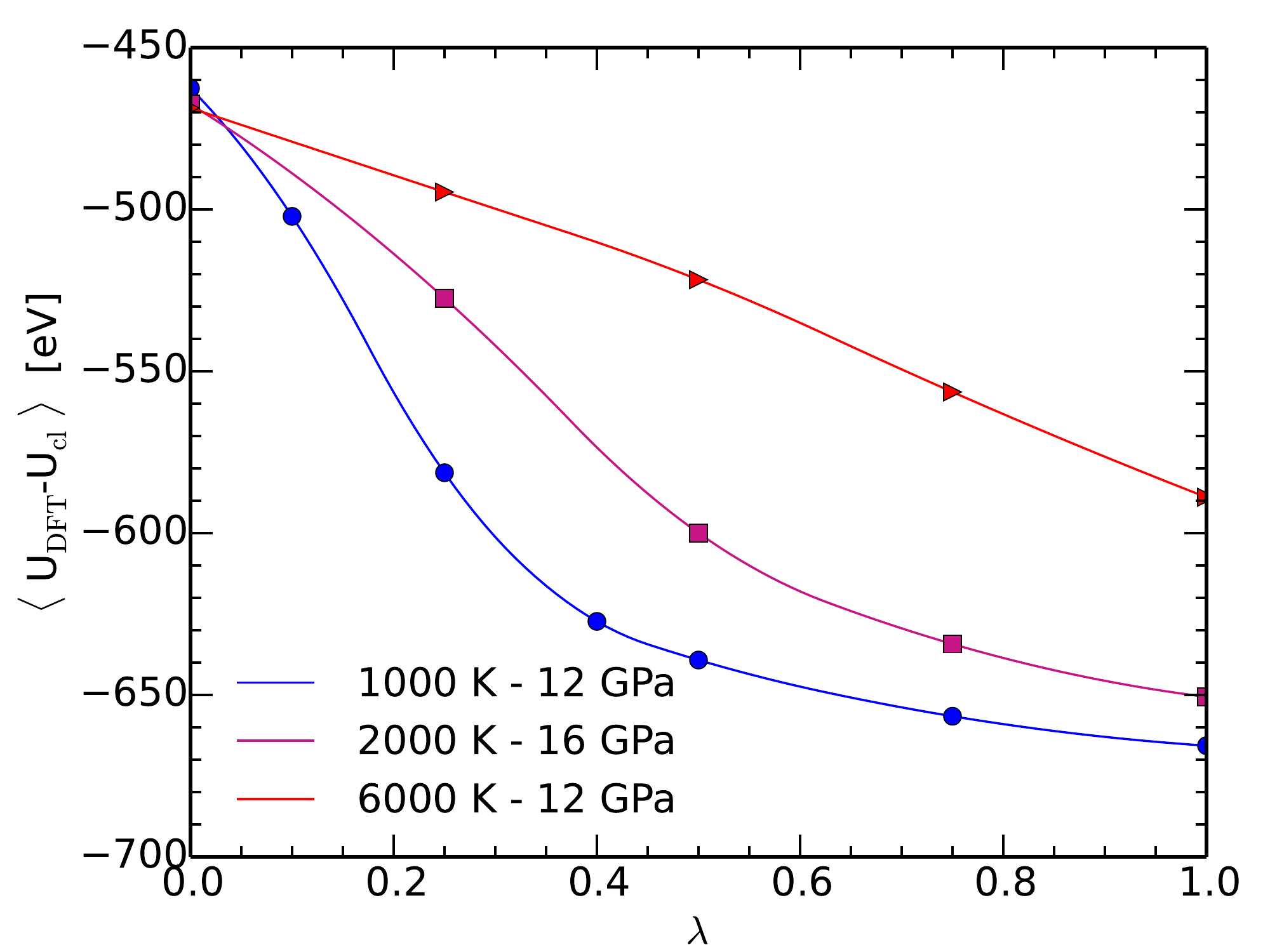}
\caption{\label{fig:vksvcl} $\langle U\indice{DFT}-U\indice{cl} \rangle$ vs $\lambda$ for 
three different $P$-$T$ conditions for a 
mixture ratio $N\indwater$:$N\indih$=24:48.}
\end{figure}

While pair potentials provide a sufficiently good description of water
at high pressure for our thermodynamic integration technique to work
well~\citep{WilsonMilitzer2012,wilson_2013}, special care needs
to be taken to describe molecular hydrogen. Pair potentials have a deep
minimum that represents the intramolecular binding. If such attractive
pair potentials are used without any repulsive many-body term,
unphysical chains and clusters of hydrogen nuclei form. In
\citet{militzer_2013}, a repulsive many-body potential was constructed
and a stable thermodynamic integration technique was obtained for
molecular and partially dissociated hydrogen. Since here we are
dealing with water-hydrogen mixtures that lead to a more diverse set
of short-lived chemical species, we pursued a different and simpler
approach. We constructed sets of {\it non-bonding} pair potentials that
we have developed and tested recently~\citep{wahl_2015}. By removing
all attractive parts from the pair potentials, we prevented the
formation of unphysical clusters. Hydrogen molecules still form
gradually as we switch from the non-bonding potentials to the DFT
forces. We determined that using between 5 to 7 $\lambda$ points was
still sufficient to accurately calculate the integral in \eq{eq:tdi}.
Examples of the evolution of this average as a function of $\lambda$
are shown in \fig{fig:vksvcl}.

To derive the Helmholtz free energy of the classical system,
$F\indice{cl}$, we performed another thermodynamic integration to a
reference system with known Helmholtz free energy, $F_0$. For this
paper, we used the ideal gas as reference. For this integration, we
used many more $\lambda$-steps as it only requires classical
simulations, which are $10^5$ times faster than DFT computations.

With this two-steps integration approach, we are able to derive the
Helmholtz free energy for the DFT system at different densities,
temperatures, and concentrations. Adding the $PV$ term, we obtain the
Gibbs free energy, $G\indice{DFT} = F\indice{DFT} + P\indice{DFT}V$,
where $P\indice{DFT}$ is the pressure given by the DFT calculation and
$V$ the volume of the simulation cell.

\begin{figure}[!ht]
\centering
\includegraphics[width=1.\columnwidth]{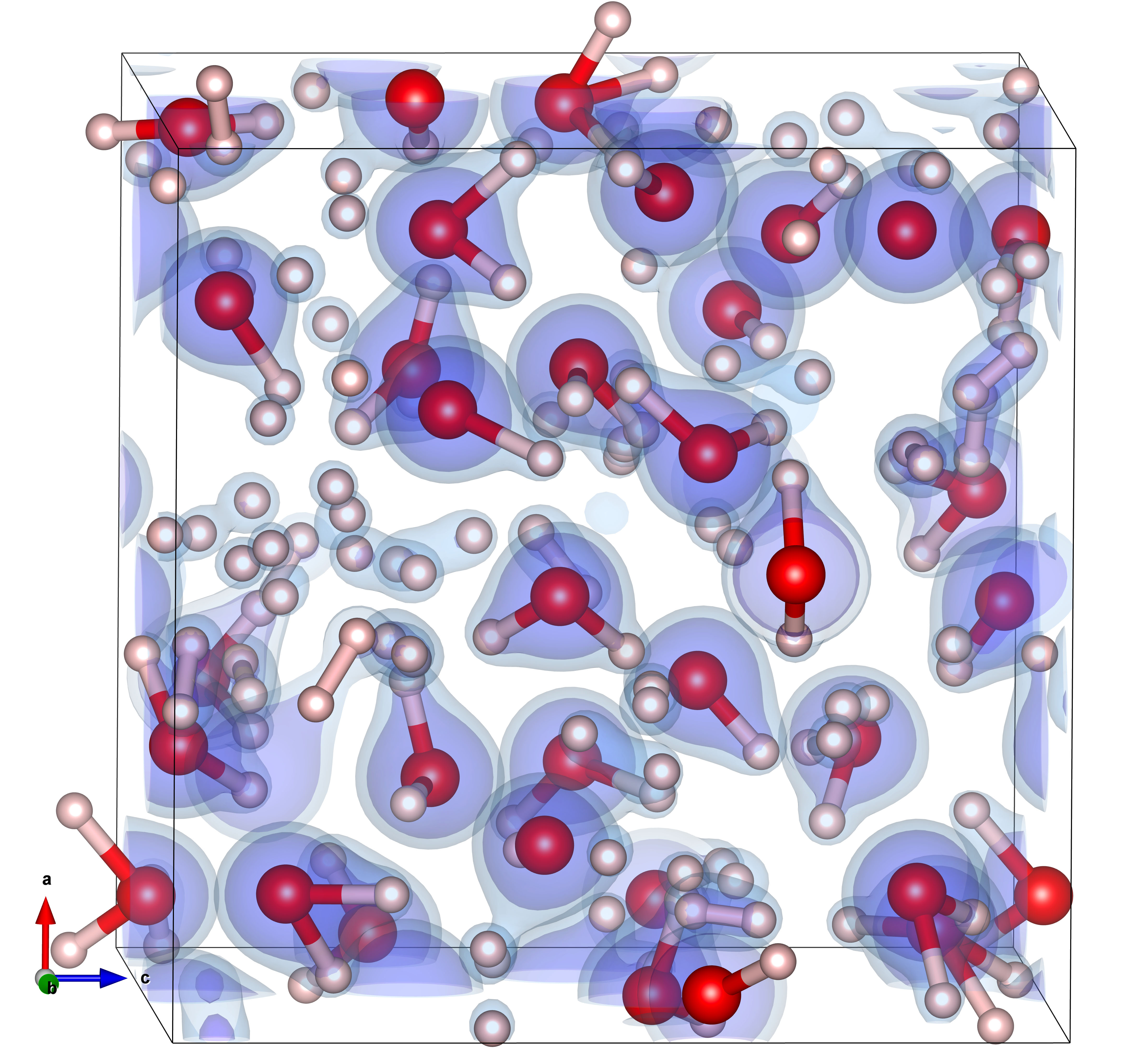}
\caption{\label{fig:snapshot} Snapshot of a simulation at 6000~K and 45~GPa for a 
$N\indwater$:$N\indih$=32:32 mixture. The 
isosurfaces show the electronic density. The bond structure is based on the nuclei 
distances.}
\end{figure}

We considered different concentrations of water molecules, $x$, 
\begin{equation} 
x = \frac{N\indwater}{N\indwater+N\indih}, 
\end{equation} 
where $N_i$ is the number of molecules of type $i$ in the simulation cell. 
We performed simulations for the following $N\indwater$:$N\indih$ ratio: 
48:0, 40:16, 32:32, 24:48, 16:64, and 0:92. Except when noted otherwise, 
the concentrations refers to the total contents in the simulation cell 
regardless what chemical species form at various pressures and 
temperatures. In \fig{fig:snapshot}, we show a snapshot from a simulation 
for $x=0.5$ and $N\indwater$:$N\indih$=32:32, which implies that 32 oxygen 
and 128 hydrogen atoms were present in the simulation cell. When we varied 
the water concentration, we always replaced one water molecule with two 
hydrogen molecules so that the volume of the simulation cell would not 
change too much. We performed simulations in pressure range from 2 to 
70~GPa along four isotherms at 1000, 1500, 2000 and 6000~K.

\section{Results}

\begin{figure}[!ht]
\centering
\includegraphics[width=\columnwidth]{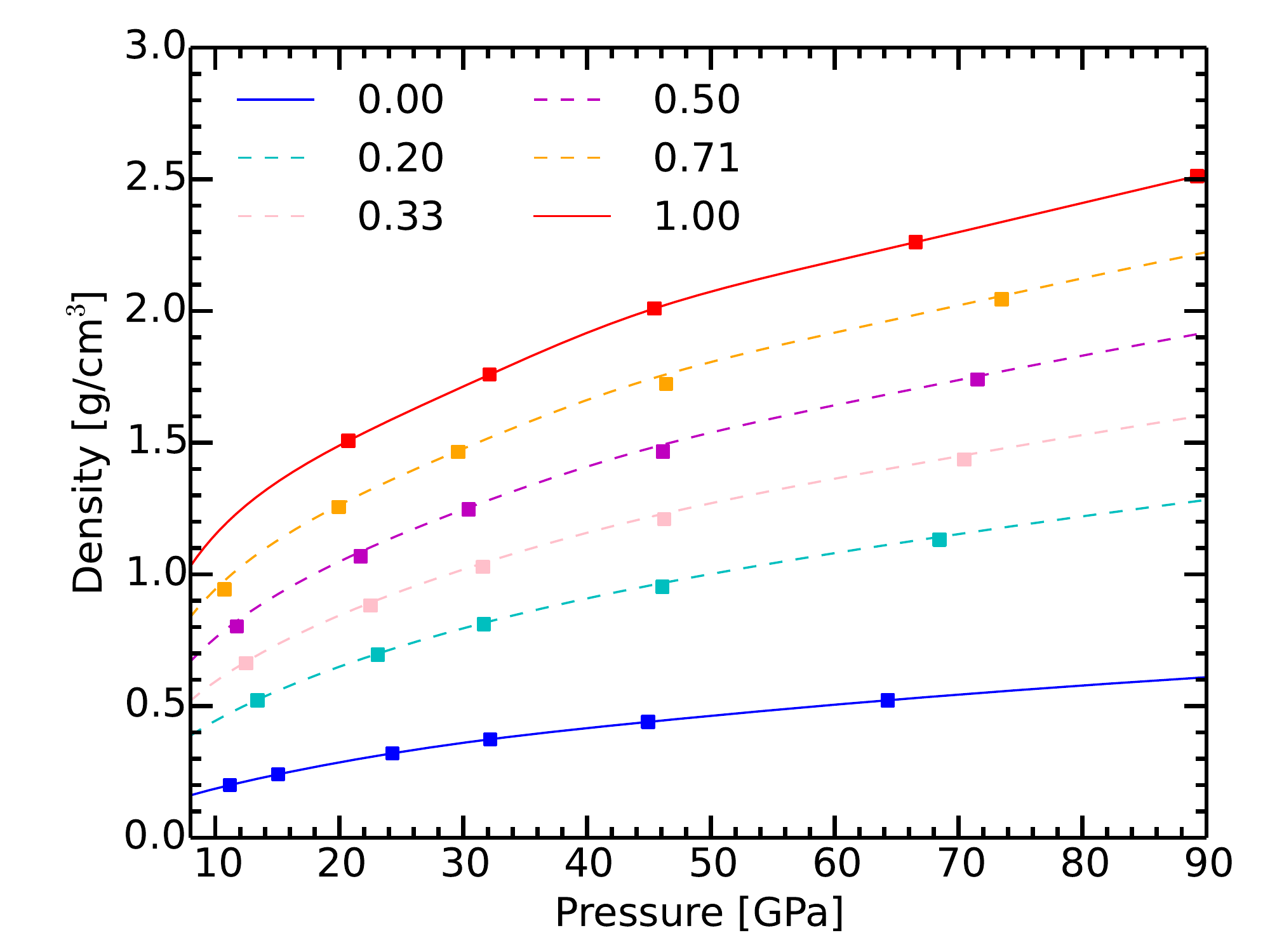}
\caption{\label{fig:density_vs_P_6000K} Density versus pressure at
  6000~K for different water concentrations, $x$, indicated in the
  legend. The squares represent our simulation results.  The full
  lines are spline interpolations for the pure systems while the
  dashed lines are the predictions from an ideal mixing approximation
  that very well reproduces our direct simulation results of mixtures.
}
\end{figure}

\begin{figure}[!ht]
\centering
\includegraphics[width=\columnwidth]{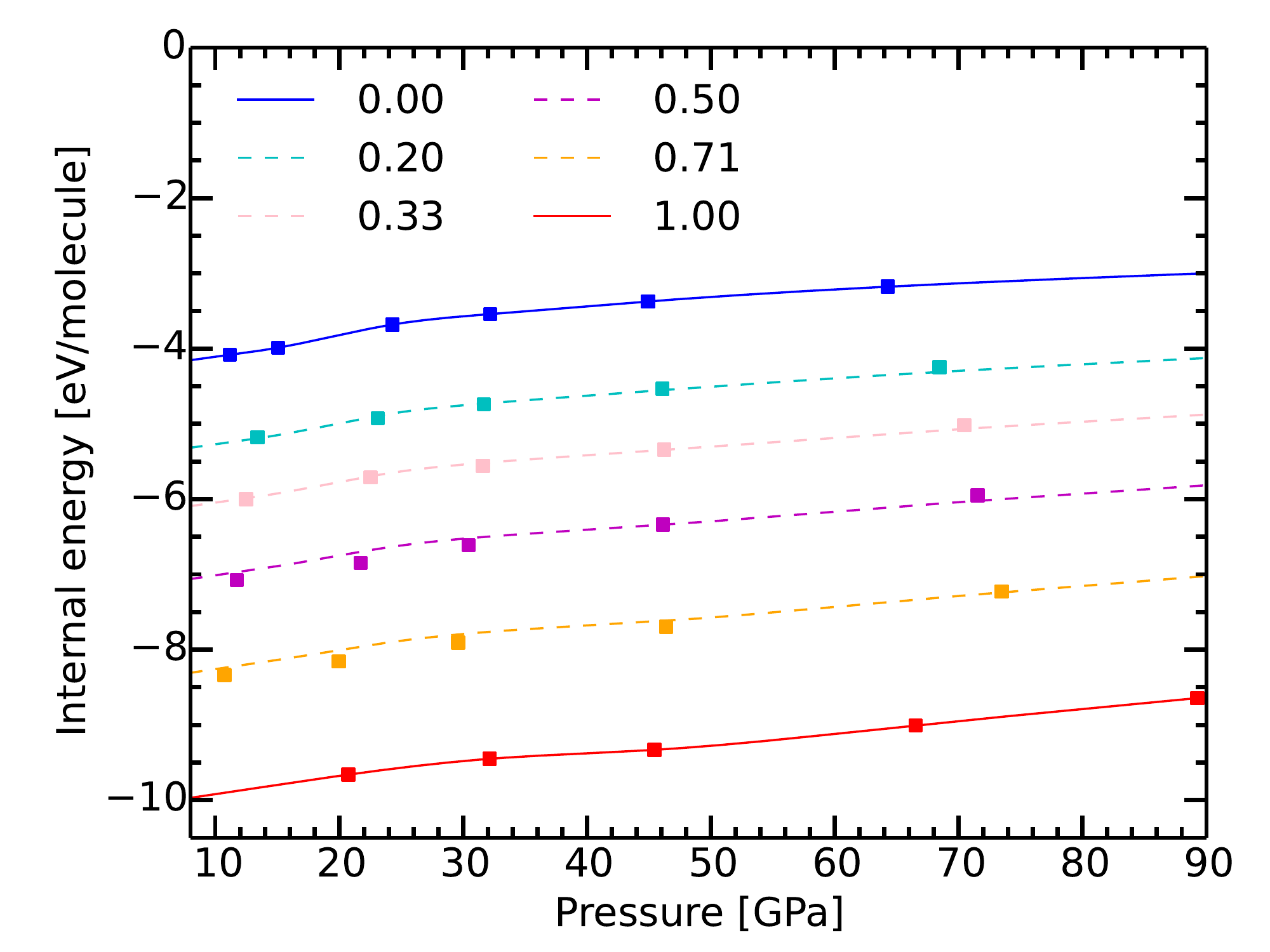}
\caption{\label{fig:E_vs_P_6000K} Internal energy versus pressure for different
concentrations $x$ reported in \fig{fig:density_vs_P_6000K}. The dashed lines are again 
the predictions from an ideal mixing 
approximation.}
\end{figure}

\begin{figure}[!ht]
\centering
\includegraphics[width=\columnwidth]{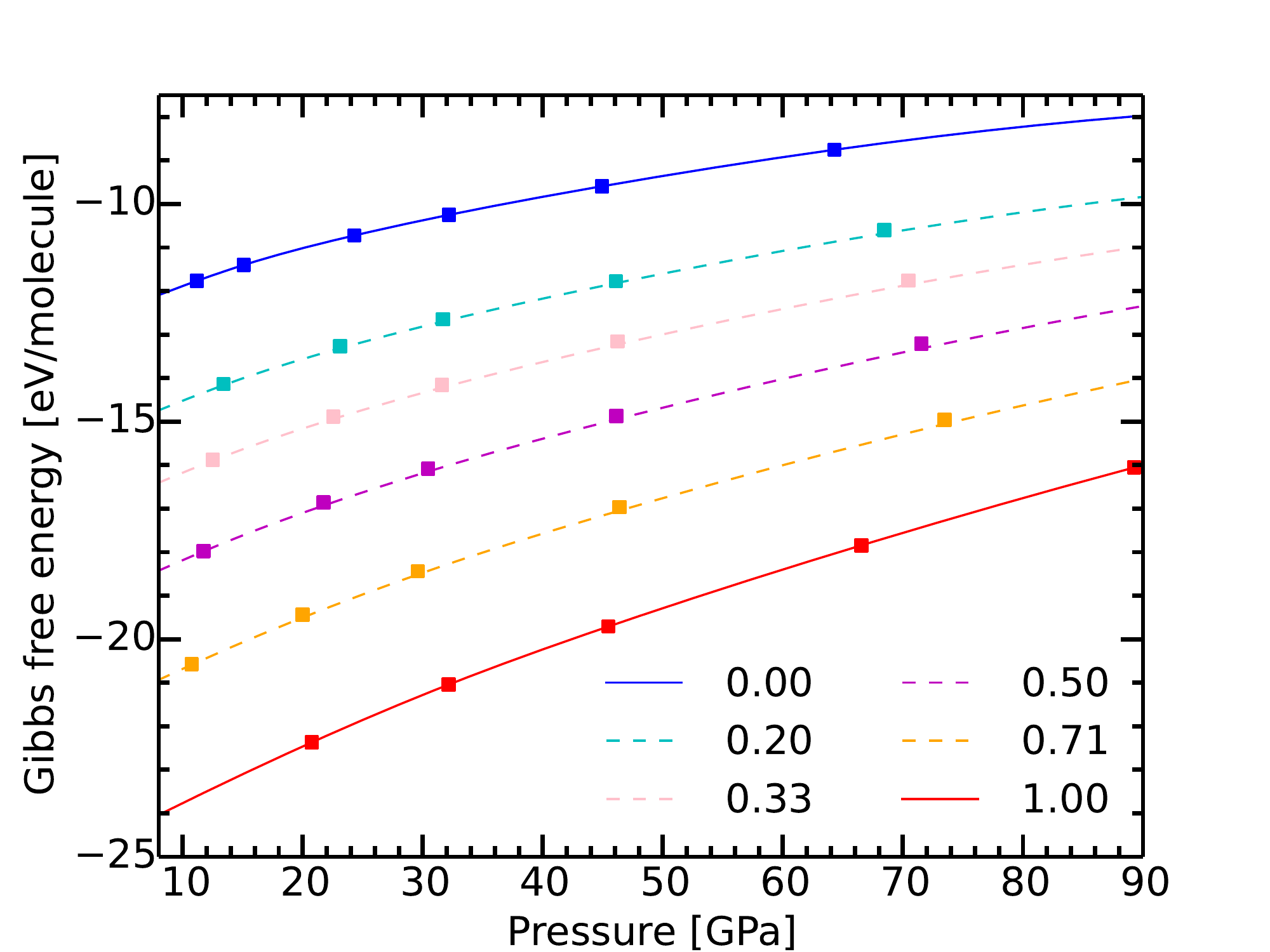}
\caption{\label{fig:G_vs_P_6000K} Gibbs free energy versus pressure for
different concentrations $x$ reported in \fig{fig:density_vs_P_6000K}.}
\end{figure}

From the DFT molecular dynamics and the thermodynamic integration, we
extracted pressure, internal energy, and Gibbs free energy as function
of temperature, density, and concentration. The results are given in
Tab.~1\footnote{The table is available in the published article in ApJ.}.

In \fig{fig:density_vs_P_6000K}-\ref{fig:G_vs_P_6000K}, we plotted
various thermodynamic quantities for different pressures and
concentrations at 6000~K . We used a spline interpolation for the pure
water and pure hydrogen system. We compared our simulation results for
the mixtures with an ideal mixing approximation using our simulation
results for H$_2$ and H$_2$O and an additive volume law at constant
pressure and temperature. The ideal mixing approximation was found to
reproduce our simulation results for the mixtures well. Some small
deviations of the order of a few percents (up to 10\% locally), in
particular for the density and for the internal energy, can be
identified, however.  Nevertheless we conclude that the ideal mixing
approximation is robust and will be sufficiently accurate for the
construction of most planetary interior models.

We computed the Gibbs free energy of mixing:
\begin{eqnarray}
 \Delta G\left(x,P,T\right) &=& G\left(x,P,T\right) -
x\,G\indwater\left(P,T\right) \nonumber\\
                            &-& \left(1-x\right)\, G\indih\left(P,T\right),
\end{eqnarray}
where $G$ is a Gibbs free energy per molecule, and $G_i$ is the Gibbs free energy of the 
pure system of molecule $i$. A 
homogeneous mixture is the thermodynamically preferred state if:
\begin{equation}
 \left.\frac{\partial^2 \Delta G}{\partial x^2}\right|_{P,T} \geq 0,
\end{equation}
which implies that $\Delta G$ is a convex function of $x$ at given
pressure and temperature. If the Gibbs free energy difference shows a
partial or a full concavity, the homogeneous system is unstable and
undergoes a partial or complete phase separation.

\begin{figure}[!ht]
\centering
\includegraphics[width=\columnwidth]{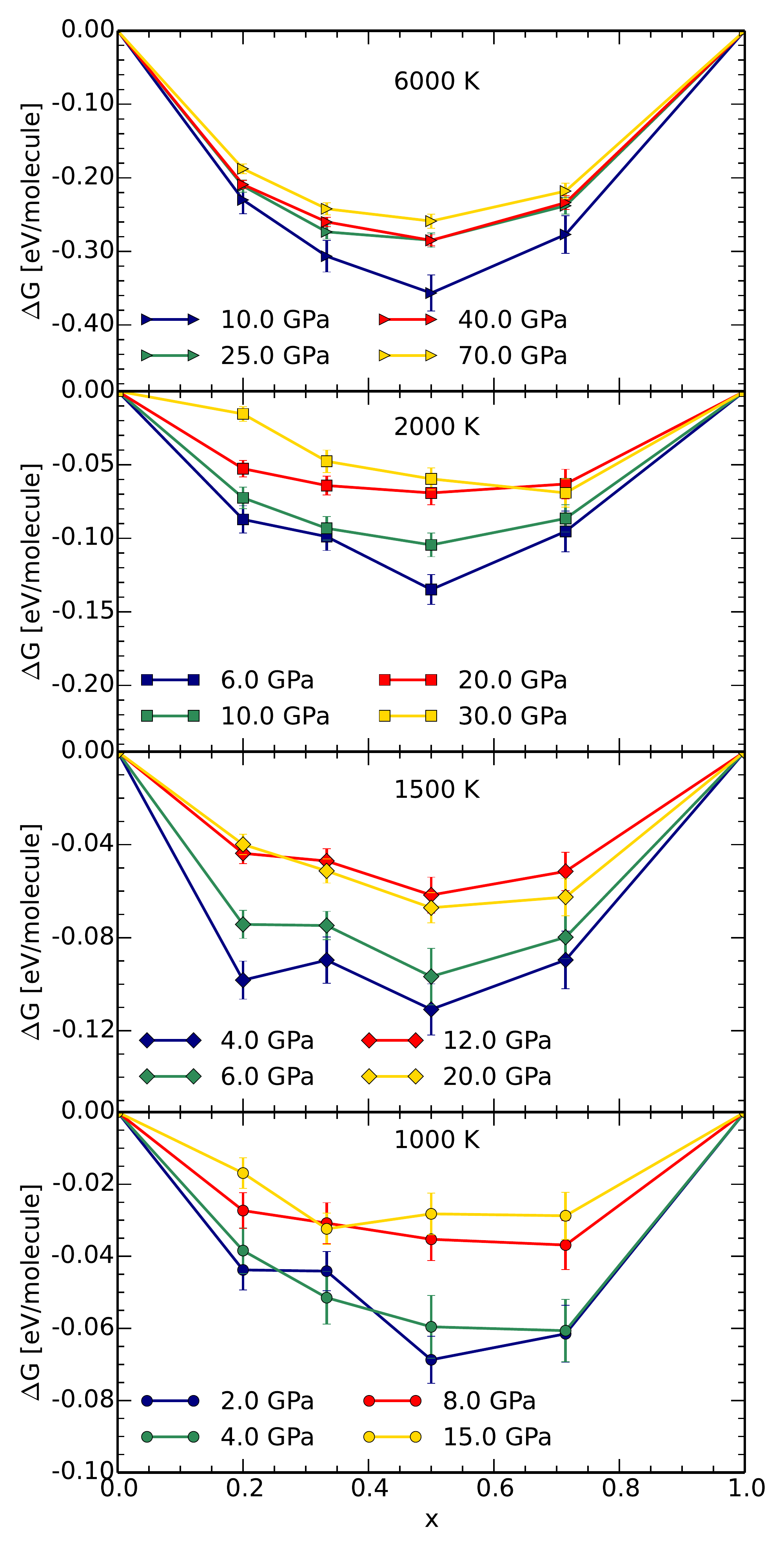}
\caption{\label{fig:hull_diag} Gibbs free energy of mixing as a
  function of the water concentration $x$ (hull diagram) at different
  temperatures and pressures.}
\end{figure}

In \fig{fig:hull_diag}, we plotted the Gibbs free energy of mixing for
different temperature and pressure conditions. Most of the curves are
convex and when they are not, a convex curve can still be drawn
within the errorbars. We therefore conclude that, for the whole set of
conditions we explored, a homogeneous mixture of hydrogen and water in
any proportion is the thermodynamically stable state. We were not able
to identify an indication for a phase separation at any condition that
we explored.

For given pressure and temperature, we can split the Gibbs free energy
of mixing into three terms, $\Delta G=\Delta E+P\Delta V-T\Delta S$.
The internal energy term, $\Delta E$, represents the interaction
between the different species in the mixture. The $P\Delta V$ term
measures deviations in density from an ideal mixture. Finally,
$-T\Delta S$ is the full entropy of mixing including ideal and
non-ideal contributions.

\begin{figure}[!ht]
\centering
\includegraphics[width=\columnwidth]{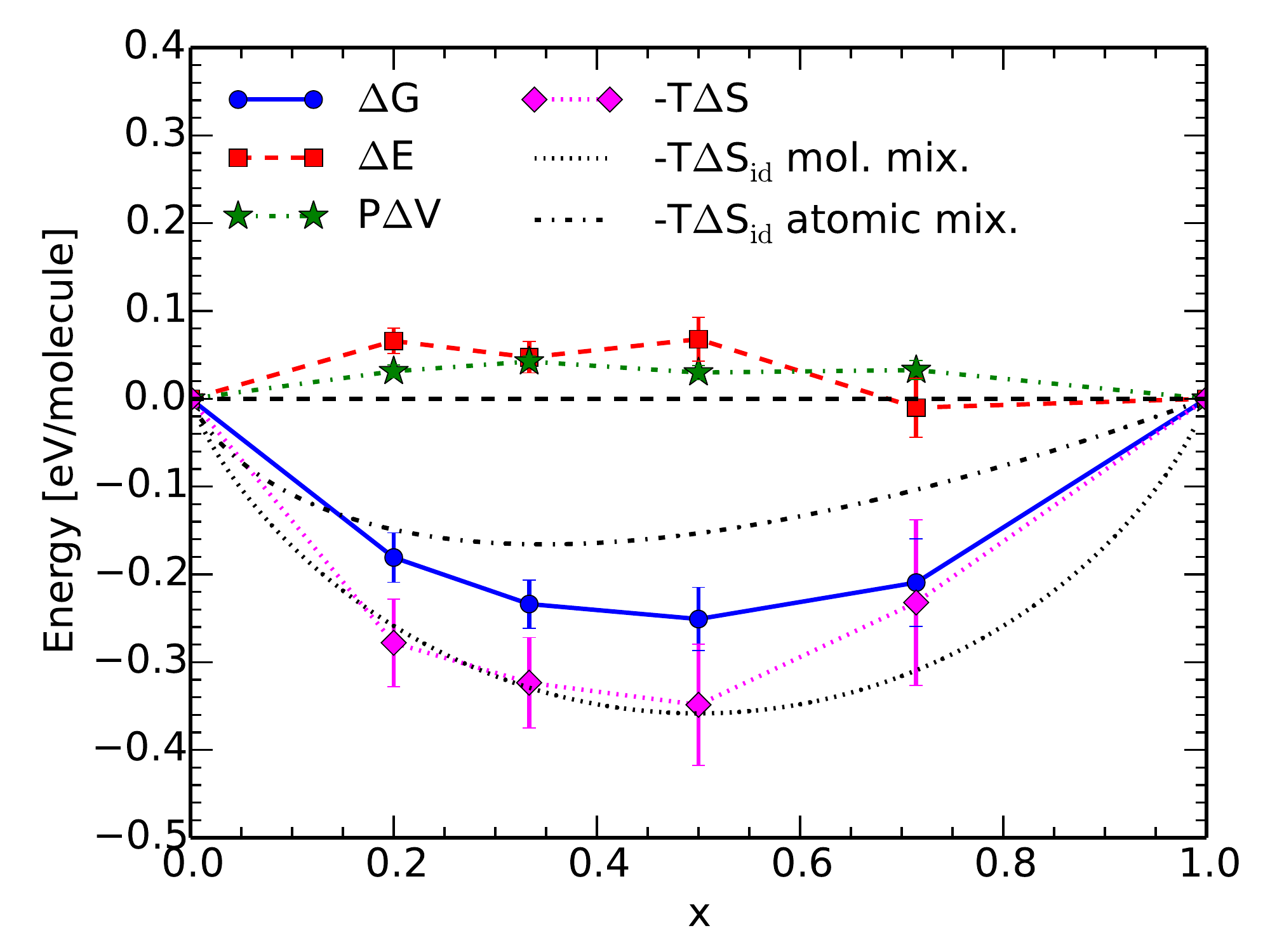}
\caption{\label{fig:differences_6000K_70GPa} Decomposition of the
  Gibbs free energy of mixing as a function of the water concentration
  $x$ at 6000~K and 70~GPa. The black dotted line shows the predicted
  contribution for an ideal entropy of mixing of hydrogen and water
  molecules while the dash-dotted line is an ideal entropy of mixing
  of oxygen and hydrogen atoms. The latter does not agree well with
  our simulation results.}
\end{figure}

\begin{figure}[!ht]
\centering
\includegraphics[width=\columnwidth]{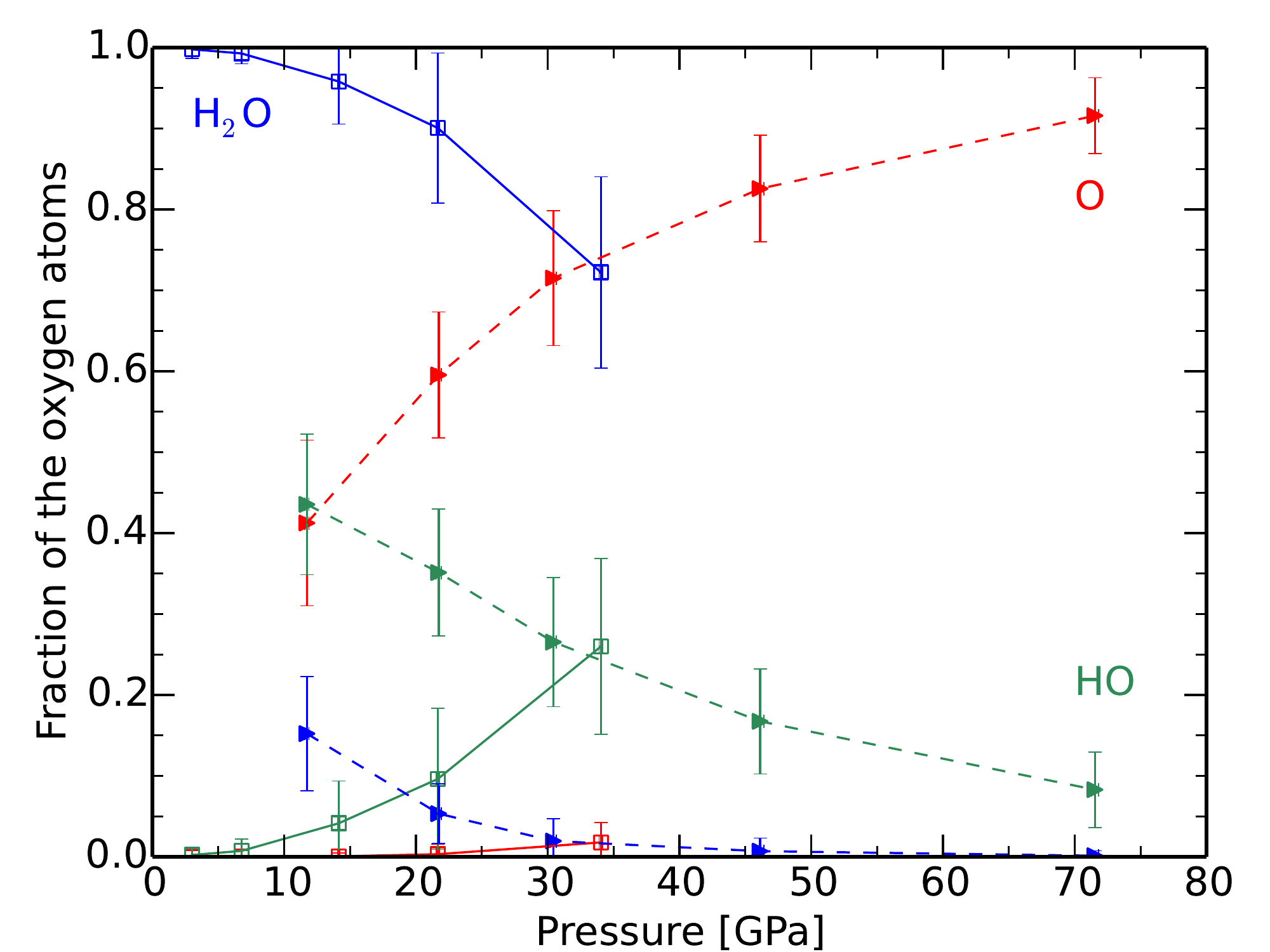}
\caption{\label{fig:dissoc_32_32_O} Distribution of the oxygen atoms among the different 
oxygen bearing species at 2000~K 
($\square$) and at 6000~K ($\blacktriangleright$) for a mixing ratio 
$N\indwater$:$N\indih$=32:32. }
\end{figure}

\begin{figure}[!ht]
\centering
\includegraphics[width=\columnwidth]{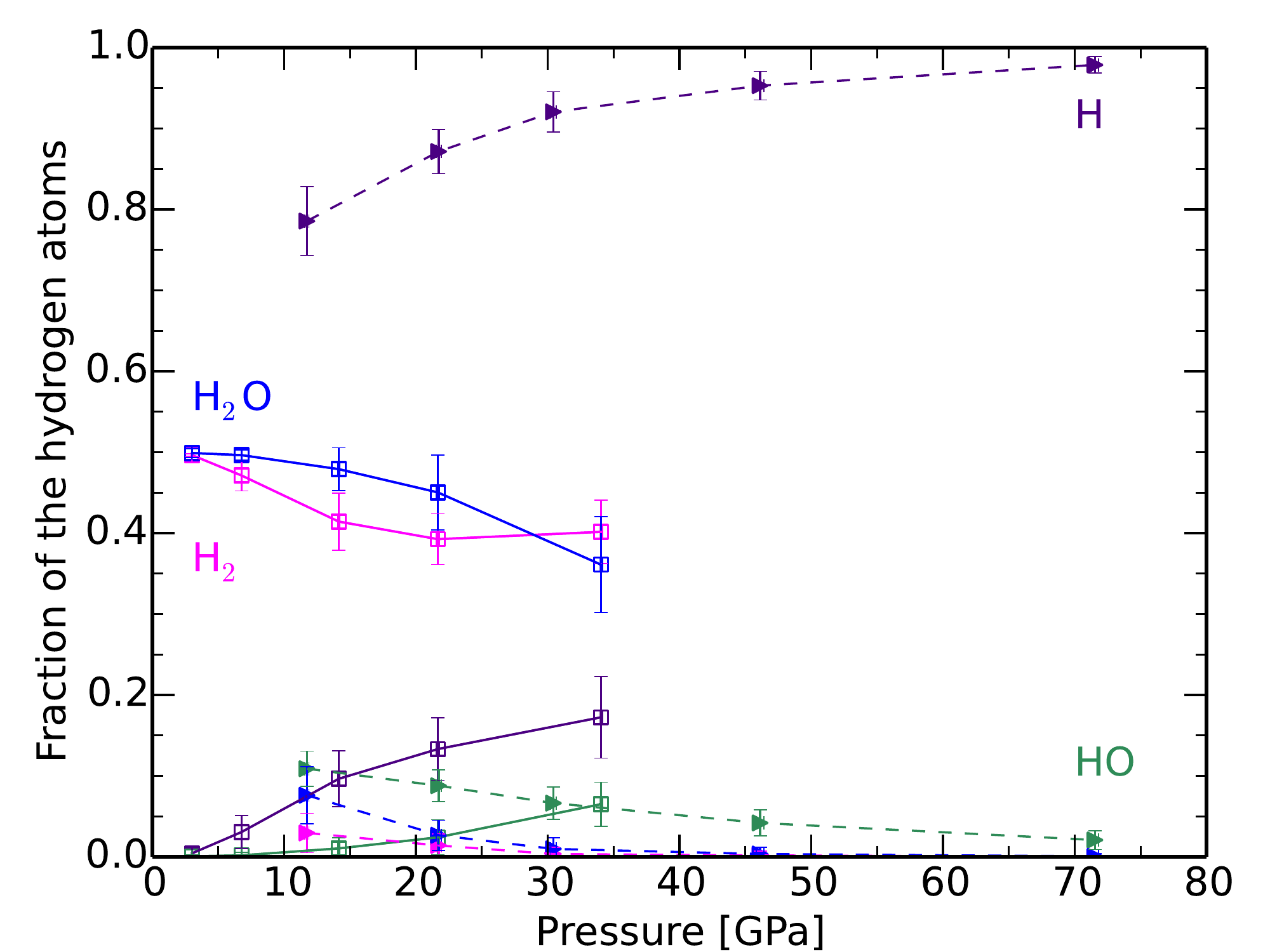}
\caption{\label{fig:dissoc_32_32_H} Distribution of the hydrogen atoms among the different 
hydrogen bearing species at 2000~K 
($\square$) and at 6000~K ($\blacktriangleright$) for a mixing ratio 
$N\indwater$:$N\indih$=32:32.}
\end{figure}

From the example at 6000~K and 70~GPa shown in
\fig{fig:differences_6000K_70GPa}, we can infer that the contributions
to the Gibbs free energy of mixing from the internal energy and the
$P\Delta V$ term are small while the entropy is the dominant term by far.
The computed entropy can be very well approximated by the ideal
entropy of mixing of water and hydrogen molecules
$S_{\rm{id}}=-k\indice{B} \left[ x\ln x + (1-x) \ln (1-x)\right]$. We
find this approximation to work very well even if the system is almost
fully dissociated at, e.g., 6000~K and 70~GPa (see
\fig{fig:dissoc_32_32_O} \& \ref{fig:dissoc_32_32_H}). If one tries to
use the entropy of mixing for two atomic systems instead, the
agreement with the simulation results is inferior
(\fig{fig:differences_6000K_70GPa}). We conclude that at high
temperature when many short-lived species are present, the
attraction between ionic species is still sufficiently large so that
molecular entropy of mixing is a much better approximation than
relying on a mixture of atoms.

We performed a systematic study of the different contributions in the
Gibbs free energy of mixing and always found that the entropy is the
dominant term that can be matched well with an ideal mixing
approximation for molecules. The largest deviations from this
approximation arise for high water concentration. The presence of
hydrogen appear to slightly alter the dissociation fraction of water
molecules in the mixtures. This is the strongest non-ideal mixing
effect that we identified. Similarly, the presence of helium atoms
appears to increase the stability of hydrogen molecules when various
hydrogen-helium mixtures are compared for given pressure and
temperature~\citep{vorberger_2007}.

In addition to studying various thermodynamic functions, we also
determined the different chemical species present in the mixture for
the different pressure, temperature, and concentration. We used a
similar species analysis method as was employed by
\citet{vorberger_2007}. At each time step, we computed the distances
between the nuclei. If a pair of nuclei remained closer than a given
distance for a minimum duration (ten times the vibration period of
molecular hydrogen $\tau\indih$=7.6~fs) we considered the two nuclei
bound. From pairs of bound nuclei, we built larger chemical species if
the bonds remained contiguous. Using the first minimum in the radial
distribution functions \citep{soubiran_2014}, the following distance
limits were derived: $l_{\scriptsize\textrm{H-H}}=1.0$~\AA,
$l_{\scriptsize\textrm{O-H}}=1.3$~\AA~and
$l_{\scriptsize\textrm{O-O}}=1.9$~\AA.

The \fig{fig:dissoc_32_32_O} and \ref{fig:dissoc_32_32_H} show
examples of the chemical compositions of the mixture. We observed that
H, H$_2$, O, HO and H$_2$O make up more than 99\% of the observed
species in the system. A few transient bigger molecules were
identified also.
The temperature has a strong impact on the degree of dissociation as
expected. While at 2000~K and below, the mixture is almost fully
molecular, at 6000~K we observe a system that is nearly fully
dissociated. We thus expect a much higher electric conductivity at
6000~K because the dissociation is generally associated with
ionization especially if hydrogen is present \citep{collins_2001}. 

For both temperatures, the fraction of dissociated species appears to
increase with pressure. Water molecules tend to split into hydroxide
and hydrogen ions or even to fully dissociate, entering into the regime
where no stable chemical bonds exist.

\section{Discussion}

The computed equation of state of the homogeneous mixtures shows only
small deviations from the ideal mixing approximation. We thus conclude
that an ideal mixing law may provide quite reasonable estimates for
the purpose of planetary modeling in general as long as no phase
separation is expected and that very accurate EOSs for the pure
systems are used.

Moreover, we do not find a phase separation for the hydrogen-water
mixtures we studied. As was observed for other mixtures at higher
pressure \citep{WilsonMilitzer2012,gonzalez_2014}, the entropy is
nearly ideal and contributes the most to the free energy of mixing,
which stabilizes the homogeneous mixture.

We have to stress here that our miscibility predictions differ
slightly from what has been inferred from high pressure experiments by
\citet{bali_2013} (see \fig{fig:phase_diag} for comparison). In these experiments, 
samples of different minerals
were compressed up to a 2.5~GPa in presence of water. Due to chemical
reactions, hydrogen was released. In some conditions a phase
separation of water and hydrogen occurred, even for temperatures above 1000~K. 
While these experiments may
be relevant for the Earth's interior, the conditions are quite
different from the envelope of the giant planets where no minerals are
present. We therefore suggest that diamond anvil cell experiments
water-hydrogen mixtures in a mineral-free environment should be
performed.

With knowledge of the experimental work by \citet{bali_2013},
  \citet{aranovich_2013} constructed a van Laar mixing model using the
  molar volume of the pure species and a van Laar parameter fitted on
  earlier experiments at the kbar regime. This model predicts
  hydrogen-water mixture to phase separate at even higher temperatures
  that suggested by \citet{bali_2013}. When we
  constructed a van Laar mixing model, we were not able to obtain a
  good fit to our \textit{ab initio} Gibbs free energies, even though
  the molar volumes in our simulations of the pure species agree quite
  well with the values that Aranovich used. To match our \textit{ab
    initio} data, the van Laar model would need to be extended by
  including nonideal mixing effects for the volume and the van Laar
  mixing parameter $W$ would need to be made temperature and pressure dependent.

The fact that our results indicate the water-hydrogen mix at 2-70 GPa
has implications on our understanding of ice giant planets. The generic
three layer model \citep{nettelmann_2013} with a hydrogen-rich outer
envelope, a water-rich intermediate layer, and a rocky core may be a
simplification. The boundary between the hydrogen and water layers is
at about 10~GPa and 2000~K in Uranus and Neptune, which falls into the
parameters regime that we explored with our simulations. Yet our
results show that such a sharp boundary between the two layers is
thermodynamically unstable. Convection will efficiently mix the two
layers unless the system approaches a semi-convective state.

Therefore we cannot rule out the three-layer picture completely
because the time scale of the mixing has to be taken into account. In
the core-accretion model~\citep{pollack_1996}, solid planetesimals are
accreted first, followed by a run-away gas accretion. The core is assumed to
differentiate quickly letting the lighter components rise to the 
surface
of the core. Based on our results, water would start to mix with
the gas but the time scale is then a key parameter. If the mixing is
only diffusive, the planet remains in a nearly differentiated state
because the diffusive time scale of water in hydrogen can be as long
as $10^{12}$ years based on the diffusion coefficient estimates in
\citet{soubiran_2014} and the size of the envelope given by
\citet{nettelmann_2013}. On the other hand, if there is a vigorous and
sustained convective activity then in a few convective time scale (on
the order of a 100 years time scale \citep[p 294]{hubbard_1984}) the
planet should be homogenized, which would be incompatible with the
multiple layers assumption. Nevertheless,
\citet{leconte_2012,leconte_2013} showed that depending on the
conditions, semi-convection may set in. If a gradient of composition
exists in the interior of a planet, gravity may prevent an efficient
convection of heavy elements. In this case, a series of alternating
diffusive and convective layers are predicted to occur. Although the
long term evolution of such a semi-convective state is not fully
understood it is a possible mechanism for maintaining fairly steep
compositional gradients.  The planet would then have a water-rich deep
inner envelope and a gradual, semi-convective transition to a
hydrogen-rich envelope.

\begin{figure}[!ht]
\centering
\includegraphics[width=\columnwidth]{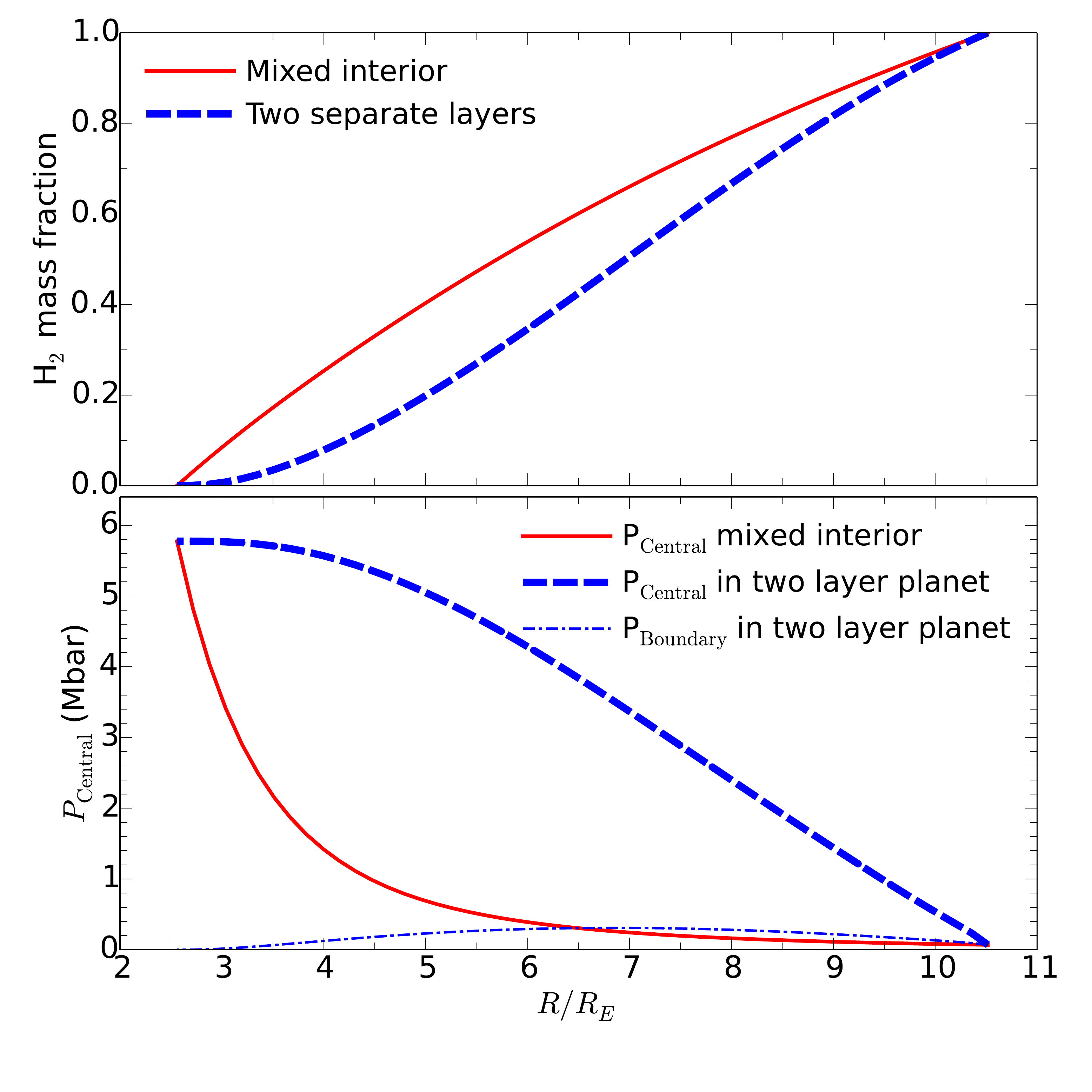}
\caption{\label{fig:MR_H2-H2O} 
Predictions from mixed and differentiated interior models for
H$_2$-H$_2$O planets with 10 Earth masses.  The upper panel shows the
inferred H$_2$ mass fraction for a given radius. The lower panel
displays the central pressure and, for differentiated planets, also
the pressure at the boundary between the two layers.}
\end{figure}

Finally, we want to discuss the importance of the water-hydrogen
miscibility for exoplanet interior modeling. Given a mass and radius
observation for a sub-Neptune planet we find the inferred hydrogen
fraction depends significantly on whether water and hydrogen occur in
mixed form. In \fig{fig:MR_H2-H2O} we compare the interior properties
of hypothetical planets composed of water and hydrogen only. While
\citet{valencia_2010} found very similar mass-radius relationships for
homogeneously mixed and differentiated, two-layer planets composed of
iron and silicates, here we find much larger deviations because
hydrogen and water have very different compressibilities. For a planet
of 10 Earth masses, we find the radius varies between 2.6 and 10.5
Earth radii for a pure water and pure hydrogen planets. Respectively,
the central pressure varies between 5.8 and 0.07 Mbar.

In \fig{fig:MR_H2-H2O}, we also compare the hydrogen fraction for fully 
mixed and differentiated water-hydrogen planets for a given radius. If a 
planet with 10 Earth masses and 4 Earth radii were detected, one would 
infer a hydrogen fraction of only 8\%~(0.8 Earth masses) if one assumed a 
differentiated interior. The central pressure would be 5.6 Mbar, and the 
pressure at the water-hydrogen boundary would be 0.12 Mbar, which is far 
below the molecular-to-metallic transition pressure in pure 
hydrogen~\citep{vorberger_2007}. On the other hand if one assumed a 
homogeneously mixed interior structure, the hydrogen mass fraction would 
increase by 25\% (2.5 Earth masses) and the central pressure would decrease 
to 1.4 Mbar, which is sufficiently high for hydrogen molecules to 
dissociate. This reaction would also increase the electrical conductivity 
of the mixture and further the generation of magnetic fields.

The inferred hydrogen fraction increases significantly if a mixed
interior is assumed because hydrogen gas is very compressible and then
some hydrogen fluid is exposed to much higher pressure. For
differentiated planets of 10 Earth masses, we find hydrogen is never
exposed to more than 0.31 Mbar regardless of the planet's radius. 
The central pressure was found to decrease with increasing hydrogen
content because water is diluted with a material of smaller density.

A difference of 0.8 and 2.5 Earth masses in hydrogen contents also has
implication for our understanding of the environment where the planet
formed. 

\section{Conclusions}

Using DFT molecular dynamics simulations, we showed that on the range
of pressure from 2 to 70~GPa and temperature from 1000 to 6000~K, the
water-hydrogen mixtures behave close to an ideal mixture. We found
that no phase separation is expected for this parameter range. A
homogeneous mixture is always thermodynamically preferred for all
concentrations. Our simulation results are in disagreement with water-hydrogen mixtures experiments in mineral cells 
\citep{bali_2013} in the lowest pressure-temperature regime explored by our calculations. 
We would suggest that high-pressure experiments of hydrogen-water mixture should be 
performed in a mineral-free environment in order to better constrain the phase diagram in this regime. 

Since we predict no phase separation for water and hydrogen, this has consequences for the ice 
giant planets. If a
planet already has a mixed water-hydrogen layer, the thermodynamic
properties, in particular the entropy of mixing, prevent any
differentiation from occurring. 

If a planet has two separate water and hydrogen layers, this implies
two things.  First, the icy materials must have been delivered early
on when very little gas was present or the icy planetesimals must have
been sufficiently large so that they penetrated through the existing
atmosphere. Both possibilities would be consistent with the core
accretion
model. 

Furthermore such a planet could not be fully convective, otherwise
water and hydrogen would mix assuming the pressure at the interface is in
the 2-70 GPa range. It is possible, however, for the planet to remain
predominantly differentiated because the interior may assume a
semi-convective state, in which water and hydrogen would not mix
efficiently. Because the long-term dynamics of this semi-convection is
not sufficiently well understood, it is difficult to model the evolution
of planetary interiors on the gigayear time scale when compositional
gradients are present.

We also showed that, unlike in the case of rock-iron mixtures,
the mixing of water and hydrogen can drastically increase the
estimates of the hydrogen content of sub-Neptune exoplanets. This
effect has to be taken into account when their interior structure and
evolution are modeled.

\section*{Acknowledgments}
\acknowledgments

This work was supported by NASA and NSF. Computers at NAS and NCCS were used.

\end{document}